\documentclass{aa}  

\usepackage{graphicx}
\usepackage{txfonts}
\usepackage[colorlinks=true,linkcolor=blue,urlcolor=blue,citecolor=blue]{hyperref}
\begin{document} 

   \title{Complex gas flows in magnetized protoplanetary disks promote the formation of dust traps at low fragmentation velocities}


   \author{Vignesh Vaikundaraman \inst{1}, 
          Joanna Dr{\k{a}}{\.z}kowska\inst{1},  
          Nerea Gurrutxaga \inst{1},
          Xue-Ning Bai \inst{2,3}}

   \institute{Max Planck Institute for Solar System Research, Justus-von-Liebig-Weg 3, 37077 Göttingen, Germany \and Institute for Advanced Study, Tsinghua University, Beijing 100084, China \and Department of Astronomy, Tsinghua University, Beijing 100084, China}

   \date{Received September XXXX; accepted YYYY}
   
 
  \abstract
   {Non-ideal magnetohydrodynamic simulations of protoplanetary disks show a plethora of complex gas structures, including winds, rings, and gaps. These affect dust transport and help form dust traps, which are essential for planetesimal formation. Although studies have explored the evolution of dust in such systems, they have done so either in 1D or without dust coagulation, and the effect of such systems on dust growth is still an active area of research. }
   {This work aims to investigate the effect of a complex gas flow architecture on global dust evolution, including dust growth and transport. We examine the timescales of different processes impacting dust evolution and discuss prospects of forming planetesimals.}
   {We post-process gas velocity output from a 2D non-ideal magnetohydrodynamic simulation using a 2D (r-z) Monte Carlo dust coagulation code to perform global simulations of dust growth and evolution. We perform three runs, one with a typical steady-state disk and two with the gas velocity from the MHD simulation, where we vary the fragmentation velocity.}
   {Our results show that the advection of small particles by the gas due to strong gas velocities can play an important role in setting the dust size distributions around protoplanetary disks. The gas flow structure has a transition region, and this region acts as a location of a dust pile-up, increasing the pebble-to-gas ratio by a factor of 2.5 when compared to the steady state disk. Lowering the fragmentation velocity improves the stability of the pile-up, but the pebble concentration is not as high. This scenario acts as a way to form a dust trap in a disk without a pressure bump. We discuss the possibilities for planetesimal formation in such a trap.}
   {}

    \authorrunning{Vaikundaraman et al.}
   \titlerunning{Dust trap formation in complex gas flows }
   \maketitle

\section{Introduction}\label{sec:introduction}
Protoplanetary disks are accretion disks transporting material to the central star. A young solar mass star has an accretion rate of  $10^{-9} - 10^{-8} \mathrm{M}_\odot/\mathrm{yr}$ \citep{Manara2023}. The transport of angular momentum governs the large-scale dynamics of a protoplanetary disk and therefore influences the processes leading to planet formation. Turbulent transport of angular momentum was thought to be the primary mechanism with magnetorotational instability \citep[MRI;][]{Balbus1991} being the leading process to generate the turbulence in disks. However, the weak ionisation levels of protoplanetary disks posed a challenge. Non-ideal magnetohydrodynamic (MHD) effects like Ohmic dissipation and ambipolar diffusion are known to reduce the coupling between gas and magnetic fields in weakly ionized gas and therefore suppress MRI in the disk \citep{Bai2013, Gressel2015}, making MRI active only in the very inner disks and in the outer disks. Furthermore, the turbulence levels observed in protoplanetary disks are not high enough to sustain the accretion rates seen in protoplanetary disks \citep{Villenave2025}. Even though the MRI is suppressed, other turbulent processes, such as the Vertical Shear Instability \citep[VSI;][]{Nelson2013} and Convective Overstability \citep{Lehmann2025}, are still theorized to be active in these regions of the protoplanetary disks.

In recent years, an alternative paradigm for angular momentum transport through magnetized winds has been well established as a viable mechanism \citep{Blandford1982, Bai2016}. Especially in regions where MRI turbulence is suppressed, winds can be an efficient way of transporting angular momentum \citep{Bai2017}. More observations are detecting winds in disks with ALMA and JWST \citep{Whelan2023,Pascucci2023,Duchene2024}, although it is a challenge in trying to disentangle the degeneracy whether they are photoevaporative winds or MHD winds \citep{Fang2023, Rab2023, Weber2025}. The observed disk sizes can also be explained by an MHD wind-driven accretion paradigm with the help of semi-analytical disk wind models \citep{Trapman2022, Tabone2025}, thereby stressing the importance of including disk winds into all facets of protoplanetary disk evolution and planet formation. To understand and incorporate magnetized winds, it is of prime importance that we understand the MHD of disks.
 
Simulations of disks with magnetic fields have become more realistic with the inclusion of detailed physics of non-ideal MHD effects \citep{Lesur2023}. A common aspect of these simulations is the prominent presence of self-organising gas flows in the disk. The first kind of flows are zonal flows \citep{Riols2019, Bethune2020, Nolan2023, Suriano2017, Suriano2018, Suriano2019}, where self-organising gas rings are observed. \citet{Riols2019} proposed that the rings arise due to an instability that removes more mass vertically in regions with a large concentration of vertical magnetic flux. \citet{Suriano2018,Suriano2019} suggested that these structures arise due to reconnection events of the poloidal field lines. The second class of flows is where the vertical symmetry of the flow is broken, leading to a complex gas flow system in the disks. This happens due to the amplification of the toroidal magnetic field by the Hall Shear Instability  \citep[HSI;][]{Kunz2008,Bai2017, Mori2025, Sarafidou2024}. For such a system to occur, the polarity of magnetic fields becomes important to ensure the effectiveness of the Hall effect in the disk. Irrespective of the kind of gas flows that are formed, it is clear that the magnetic fields threading the disk, the gas, and the physics of their coupling (non-ideal MHD effects) are driving such gas flows. 

Dust and gas have interwoven lives in disks, and the gas flow configuration is going to play a role in how dust is transported across the disk. \citet{Nolan2023, Riols2020, XiaoHu2022,Hsu2025} suggested that these flows could lead to dust traps and hence potentially create sites for planetesimal formation. However, the simulations included dust as a pressureless fluid with fixed grain size. This is due to the fact that combining (magneto-)hydrodynamics and dust growth self-consistently is a challenge. New methods such as TriPoD \citep{Pfeil2024} are being developed to better model dust coagulation in hydrodynamic simulations. Particle-based simulations offer the possibility to post-process MHD simulations to perform dust transport/growth simulations \citep[e.g, ][]{VanClepper2025}. \citet{Hu2021} performed local 1D column dust transport simulations and saw that the flow structures created by the HSI changed the way dust particles of different sizes were transported across the disk due to the flow self-organisation. In some cases, these flows contributed significantly to the dust diffusion, increasing the effective diffusion $\alpha$ to $10^{-2}$ even in a system with a low turbulence of $\alpha \sim 10^{-4}$. We extend this approach, where we post-process results of MHD simulations with a 2D global dust coagulation code, so the effect of the gas is modelled not just on dust transport but also on dust growth. \citet{Drazkowska2018} used a similar method to understand the dust evolution in the circumplanetary disk. With recent ALMA observations being able to resolve the multi-dimensional velocity structures in protoplanetary disks \citep{Teague2019,Pinte2025}, we must investigate the effect of such flows on dust growth and planet formation.

 Forming planetesimals in disks requires a high dust-to-gas ratio, much higher than the initial values \citep{Birnstiel2024}.  Planets are efficient in creating pressure traps \citep{Rice2006}, which can efficiently form planetesimals \citep{Lau2022}. But this requires a planet to already be present in the disk to create efficient pressure traps. If the complex gas flows can create dust traps, they pose a promising solution to the formation of dust traps without having to invoke an already formed planet. The goal of the paper is to understand the effects of the gas dynamics enforced by the magnetic fields on dust growth on a global scale and its implications on planetesimal formation. In this paper, we post-process gas velocity data from a non-ideal MHD simulation detailed in \citet[][hereafter B17]{Bai2017} and perform 2D (r-z) global dust coagulation simulations. The paper is organized as follows. In Sect. \ref{sec:methods}, we describe the disk model and the simulation setups that we use to perform our simulation. In Sect. ~\ref{sec:diagnostics}, we describe the relevant timescales and growth barriers that we use to analyze our simulations. In Sect.~\ref{sec:results}, we present the results obtained with the gas velocities from the MHD simulations and compare them to a simulation representing the turbulent transport paradigm. In Sect.~\ref{subsec:planetesimalformation}, we discuss the implications of such flows on planetesimal formation. We discuss the implications of the results and caveats of our methods in Sect.~\ref{sec:discussion}. Section~\ref{sec:summary} summarizes our work.

\section{Methods}\label{sec:methods}
We use \texttt{mcdust}, a 2D(r-z) Monte Carlo dust coagulation code detailed in \citet{Drazkowska2013} and \citet{Vaikundaraman2025b}. The code is written in \texttt{FORTRAN 90} and parallelized using \texttt{OpenMP}. The code uses a representative particle approach to simulate dust coagulation \citep{Zsom2008}. Instead of following every physical particle, we track a limited number of representative particles and the collisions they go through. The fundamental working of the dust evolution aspects of the code is the same as in \citet{Drazkowska2013}. The code models dust collisional growth, fragmentation, radial drift, advection driven by gas flows, vertical settling, and turbulent mixing in vertical and radial directions. We also include erosion as a collisional outcome, where a small particle can chip a piece of a large particle if the large particle has a mass much higher than the small particle \citep{Guettler2010}. The threshold for erosion is given by the erosion mass ratio, as defined by 
\cite{Stammler2022}, and we set it to a default value of 10. 
\subsection{Disk structure}
We consider a power law disk model with the gas surface density
\begin{equation}
    \Sigma_\textrm{g}(r) = \Sigma_0 \left(\frac{r}{\mathrm{AU}}\right)^{-p},
\end{equation}
where $\Sigma_0=800$~g/cm$^{-2}$  is the gas surface density at $1 \;\mathrm{AU}$ and $p=1$. We assume a vertically isothermal disk, and the temperature is also modelled by a power law profile given by
\begin{equation}
    T(r) = T_0 \left(\frac{r}{\mathrm{AU}}\right)^{-q},
\end{equation}
with $T_0=200$~K being the temperature at $1 \;\mathrm{AU}$ and $q=0.5$. 
We assume the gas to be in local hydrostatic equilibrium, and the gas vertical density profile in that case is given by
\begin{equation}
    \rho_g(z) = \frac{\Sigma_g}{\sqrt{2\pi}H_g}\mathrm{exp}\left(\frac{-z^2}{2H_g^2}\right),
\end{equation}
where $H_g=c_s/\Omega_k$ is the gas pressure scale height, with $c_s$ being the sound speed and $\Omega_k$ being the Keplerian frequency.

\subsection{Gas and dust dynamics}\label{sec:gasdustdyn}

\begin{figure}
   \centering
\includegraphics[scale=0.6]{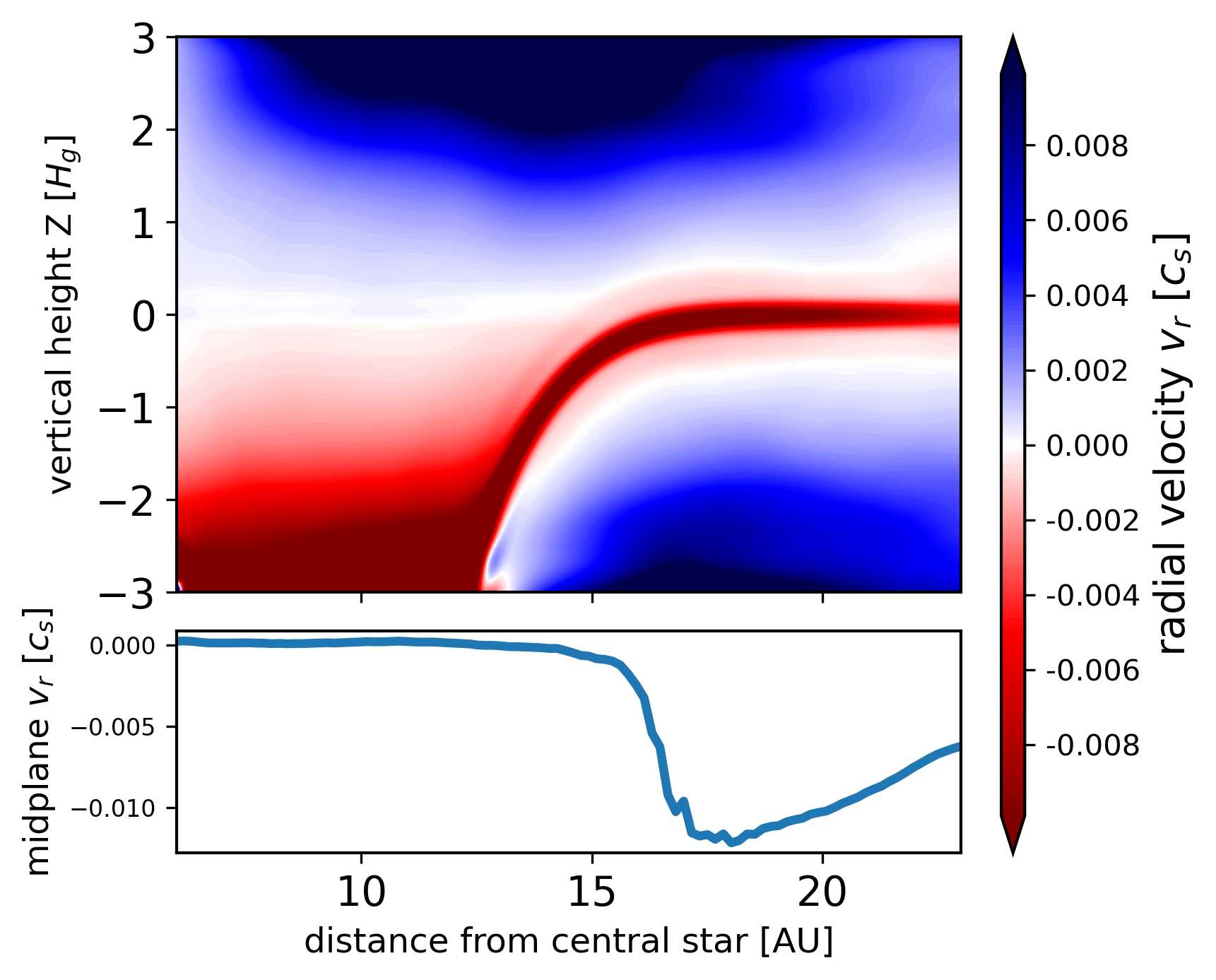}
   \caption{(Top) The 2D spatial structure of the radial gas velocity from the non-ideal MHD simulation performed by \citet{Bai2017}. (Bottom) The midplane radial gas velocities for the same gas velocity data.}
    \label{fig:windvel}
    \end{figure}

The strength of the coupling between gas and dust is characterized by the Stokes number
\begin{equation}
    \mathrm{St} = \Omega_k t_s,
\end{equation}
where $t_s$ is the stopping time of the dust particle in the surrounding gas.

The radial dust velocity (excluding the turbulent diffusion) is given by
\begin{equation}
    v_{d,r} = \frac{v_g}{1+\mathrm{St}^2} + \frac{2v_n}{\mathrm{St} + \mathrm{St}^{-1}},
\end{equation}
where $v_n = \partial_r P_\mathrm{g}/2\rho_g\Omega_k$ is the maximum drift velocity, which depends on the gas pressure gradient $\partial_r P_\mathrm{g}$. $v_g$ is the gas velocity, which, depending on the simulation setup, is taken as an input from the non-ideal MHD simulation, or we use the steady state gas velocity that describes a steady state gas in a viscously evolving disk. We discuss the details of the velocities in the following sections.

\subsubsection{Steady-state gas velocity}\label{subsubsec:ssgas}
We compare our MHD gas velocity model to the classical model of a viscous disk. To describe the gas velocity of a steady-state viscous disk, we use
\begin{equation}\label{eq:steadygasvel}
    v_g = - \frac{3\nu}{2r},
\end{equation}
where $\nu$ is the kinematic viscosity defined by
\begin{equation}
    \nu = \alpha c_s H_g,
\end{equation}
with $\alpha$ being the parameterized turbulence strength of the disk \citep{Shakura1973}.

\subsubsection{MHD gas velocity}\label{subsubsec:mhdgas}

We take the radial gas velocity profiles from the Fid+ simulation detailed in B17 where the net poloidal magnetic field is aligned with the disk rotation. We summarize the features of the simulation here and refer the reader to B17 and \citet{Hu2021} for the detailed description. The Hall Shear Instability (HSI) amplifies the midplane toroidal magnetic field, thereby creating a gas flow architecture as shown in Fig. \ref{fig:windvel}. The outer region has a symmetric flow structure where the midplane has an enhanced accretion flow, and the regions above and below the midplane have flows directed away from the central star. At $\sim 15~\mathrm{AU}$, there is a transition region from the symmetric flow structure to an antisymmetric flow structure in the inner region ($r=6-15~\mathrm{AU}$) where the region below the midplane has an enhanced accretion flow towards the central star and the region above the midplane has a flow directed away from the central star. This flow is largely laminar, and we time-average the data to make sure the small-scale numerical effects are avoided. It has been shown that turbulence (e.g, VSI in the dead-zone) can co-exist with such large-scale flows in magnetized disks \citep{Cui2020}. Therefore, the dust dynamics are also affected by turbulent diffusion, and we assume $\alpha=3\cdot10^{-4}$ to model it \citep{Jiang2024, Villenave2025}. We refer the reader to \citet{Drazkowska2013} (their Eqs. 13-18) for more details on the implementation of the diffusion of dust particles due to turbulence.

\subsection{Simulation setup}\label{sec:simsetup}
We setup a 2D (r-z) global simulation with the inner boundary at $r_\mathrm{in} = 6~\mathrm{AU}$ and the outer boundary $r_{\mathrm{out}} = 23~\mathrm{AU}$. We primarily perform two kinds of simulations, the \texttt{fidmhd} simulation, where the gas velocities from B17 are used and the \texttt{fidss} simulation, where we use the steady state gas velocities described in Sect.~\ref{subsubsec:ssgas}. The parameters of the simulations are described in Table \ref{tab:fidparams}. We initiate $\sim 10^6$ representative particles throughout the simulation domain to model the effect of dust growth and dynamics in the disk. We set the initial size of dust grains to one micron. We also perform a simulation \texttt{mhdlowvfrag}, where we lower the fragmentation threshold velocity. The list of simulations and the differences in their setups are summarized in Table~\ref{tab:listofsims}.

\begin{table}
    \centering
     \caption{The parameters and their values used in the simulations.}
    \begin{tabular}{c c}
    \hline
    Parameter & value \\
    \hline
    \hline
         dust-to-gas ratio $Z$ & 0.01 \\
         turbulence parameter $\alpha$ & $3\cdot 10^{-4}$ \\
         Gas surface density at 1 AU $\Sigma_0$ & 800 g/cm2 \\
         Temperature at 1 AU $T_0$ & 200 K \\
         Surface density power law exponent $p$ & 1 \\
         Temperature power law exponent $q$ & 0.5 \\
         material density $\rho_s$ & 1.2 g/cm3 \\
         erosion mass ratio & 10 \\
         Inner radial boundary $r_{\mathrm{in}}$ & 6 AU\\
         Outer radial boundary $r_{\mathrm{out}}$ & 23 AU\\
         monomer size $a_0$ & 1 $\mu m$\\
         Number of radial cells $N_r$ & 128 \\
         Number of vertical cells $N_z$ & 64 \\
         Number of particles per cell $N_{\mathrm{cell}}$ & 256 \\
    \hline
    \end{tabular}
   
    \label{tab:fidparams}
\end{table}

\begin{table*}
    \centering
    \begin{tabular}{ccc}
    \hline
        Run  &  gas velocity $v_g$ & fragmentation velocity $v_{\mathrm{frag}}$ \\
    \hline
    \hline
        \texttt{fidmhd}  & B17 (Fig. \ref{fig:windvel})  & 250 cm/s \\
        \texttt{fidss} & steady state velocity (Eq. \ref{eq:steadygasvel}) & 250 cm/s \\
        \texttt{mhdlowvfrag} & B17 (Fig. \ref{fig:windvel}) & 100 cm/s \\
    \hline
    \end{tabular}
    \caption{The list of simulations that we perform with the gas velocity prescription and fragmentation velocity used for each simulation.}
    \label{tab:listofsims}
\end{table*}
Dust disks of protoplanetary disks are typically larger than the outer boundary of our simulation domain, $r_{\mathrm{out}} = 23~\mathrm{AU}$ \citep{Vioque2025}. This means that the dust present outside of the computational domain can collide, grow, and drift into the simulation domain. Given that the typical lifetime of a dust disk is between $10^{5} - 10^{6}$ years \citep{Drazkowska2023}, the dust flux entering the simulated region at $r_{\mathrm{out}}$ is not expected to decay significantly over our simulated time of a few times $10^4$ years. To cover this, we introduce a constant dust flux by setting up a feeding column of dust particles at the outer boundary of the simulation domain. These particles can collide, grow, get transported, and enter the simulation domain through the outer boundary. Each time a particle exits the vertical column to enter the simulation domain, the particle is re-initiated at the outer boundary of the vertical column, which can again go through the cycle of collision, growth, and transport. This helps us mimic the feeding flux from the outer part of the disk, including the dust size distributions and the vertical structure.

\section{Diagnostics}\label{sec:diagnostics}
We simulate the global evolution of the dust particles up to $4 \times 10^4$ yr, and in order to characterize the dust size distributions, we compute certain quantities that can describe the underlying processes in the disk.
\subsection{Timescales}\label{subsec:Timescales}

We model the growth of dust particles, and the relevant timescale for growth from a particle of size $a_0$ to a particle of size $a_1$ is given by \citep{Birnstiel2012}
\begin{equation}\label{eq:growthtimescale1}
    t_{\mathrm{grow}} = \tau_{\mathrm{grow}}\mathrm{ln}\left(\frac{a_1}{a_0}\right),
\end{equation}
where $\tau_{\mathrm{grow}}$ is the timescale for a particle to double its size, which is given by
\begin{equation}\label{eq:growthtimescale2}
    \tau_{\mathrm{grow}} = \frac{1}{Z\Omega_K},
\end{equation}
where $Z$ is the vertically integrated dust-to-gas ratio of the disk. 

The radial drift timescale is given by
\begin{equation}\label{eq:drifttimescale}
    t_{\mathrm{drift}} = \frac{r}{|v_D|},
\end{equation}
where $v_D$ is the radial drift velocity given by
\begin{equation}\label{eq:driftvel}
    v_D = \frac{2v_n \mathrm{St}}{1+\mathrm{St}^2}.
\end{equation}

The background gas velocities are strong, as shown in Fig~\ref{fig:windvel}, and this means that the dust particles can be affected by advection from the gas. This can be described by the advection timescale given by
\begin{equation}\label{eq:advtimescale}
    t_{\mathrm{adv}} = \frac{r}{|v_{\mathrm{adv}}|},
\end{equation}
where $v_{\mathrm{adv}}$ is the advection velocity which is given by
\begin{equation}\label{eq:advvelocity}
    v_{\mathrm{adv}} = \frac{v_g}{1 + \mathrm{St}^2}.
\end{equation}
     \begin{figure*}[ht]
   \centering 
   \includegraphics{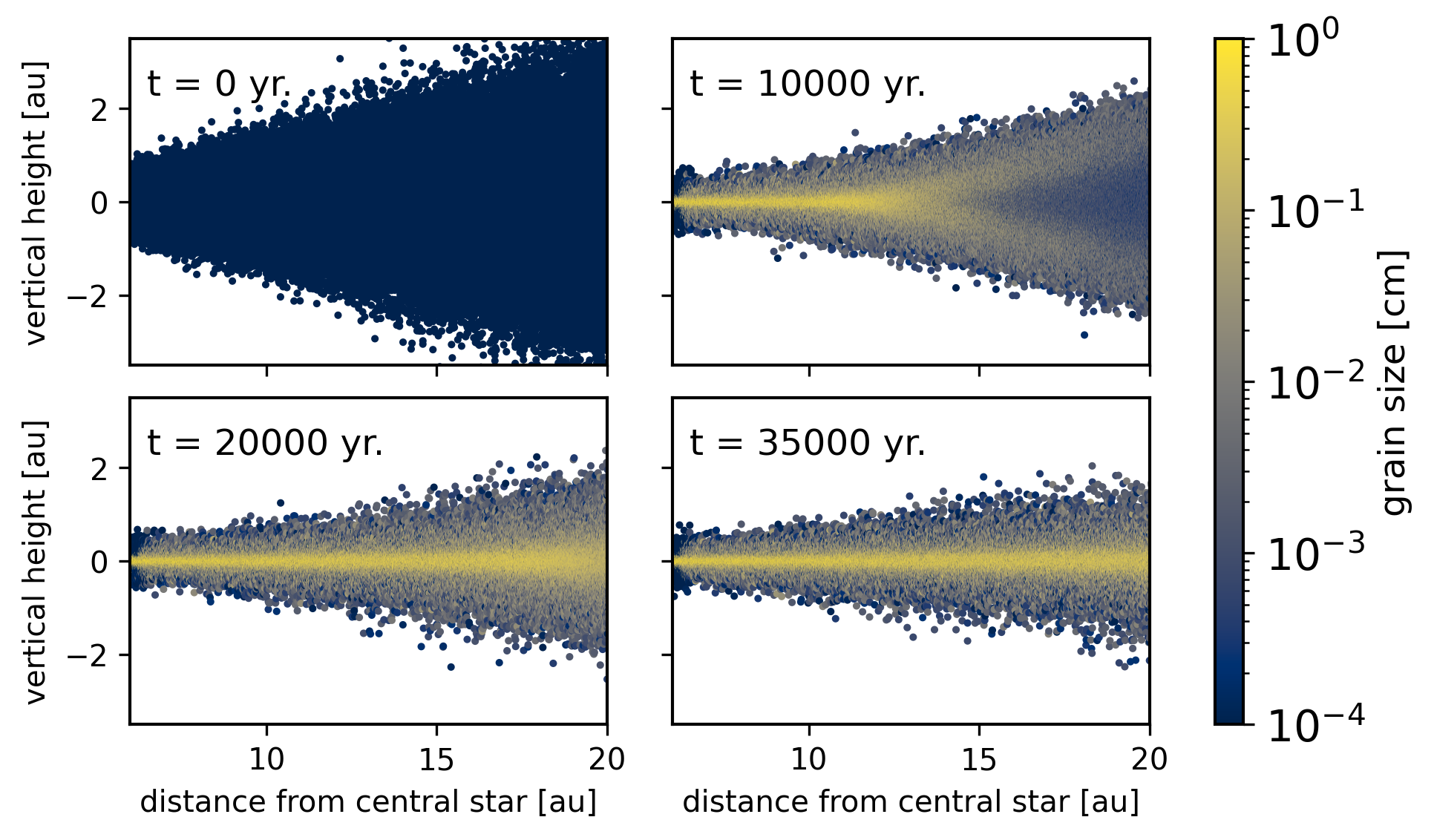}
   \caption{Scatter plot denoting the spatial distribution of the dust particles in the \texttt{fidss} simulation with the colormap showing the grain sizes of each particle. }
              \label{fig:nowindscatter}
    \end{figure*}

\subsection{Growth limits}\label{subsec:sizelimits}

Collisional velocities increase with size, and this can set a size limit to the particle growth due to fragmentation of particles after a certain size. This maximum Stokes number relating to turbulent fragmentation can be derived by equating the turbulent collisional velocity, calculated according to the formulas derived by \citet{Ormel2007}, to the fragmentation velocity. This is given by \citep{Birnstiel2012}
\begin{equation}\label{eq:frag}
    \mathrm{St}_{\mathrm{frag}} = f_\mathrm{f}\frac{v_f^2}{3\alpha c_s^2},
\end{equation}
where $v_f$ is the fragmentation threshold after which the relative velocities lead to fragmentation of particles and $f_\mathrm{f}$ is a calibration factor. 

Radial drift can set a limit to the Stokes number, and this can be derived by equating the drift timescale and the growth timescale. This is given by \citep{Drazkowska2016}
\begin{equation}\label{eq:drift}
    \mathrm{St}_{\mathrm{d}} =  f_{\mathrm{d}} \frac{2}{\pi} Z  \frac{v_K^2}{c_s^2}\left|\frac{\mathrm{d}\:\mathrm{ln}\: P_g}{\mathrm{d}\:\mathrm{ln}\: r}\right|^{-1},
\end{equation}
where $P_g$ is the midplane gas pressure in the disk.
Radial drift also contributes to the relative velocities and can lead to fragmentation of the particles. To derive a Stokes number limit, we equate the relative drift velocity between a particle of Stokes number $\mathrm{St}$ and a particle of Stokes number $N\mathrm{St}$ with the fragmentation threshold. This is given by
\begin{equation}\label{eq:driftfrag}
    \mathrm{St}_{\mathrm{df}} = \frac{v_f v_K}{ c_s^2 (1-N)}\left|\frac{\mathrm{d}\:\mathrm{ln}\: P_g}{\mathrm{d}\:\mathrm{ln}\: r}\right|^{-1}.
\end{equation}
Typically, the collisions between similar-sized bodies are the most important for mass gain/loss, and therefore $N=0.5$ is a reasonable assumption that is used to calculate the limit \citep{Birnstiel2012}.

The strong background gas flow can affect small particles that are well coupled to the gas. One can derive a Stokes number below which the advection of dust particles by the gas flow dominates over growth. This means the particles with values lower than the advection Stokes number limit are affected by the dust removal by the gas. This can be arrived at by equating the advection timescales and the growth timescale, resulting in \citep{Drazkowska2018}
\begin{equation}\label{eq:adv}
    \mathrm{St}^2_{\mathrm{adv}} = \frac{|v_g|}{Z v_K} - 1.
\end{equation}

Similar to radial drift velocities, we compute the Stokes number limit for advection-driven fragmentation, which is given by
\begin{equation}\label{eq:advfrag}
    \mathrm{St}_\mathrm{af} = \sqrt{\frac{v_f}{|v_g||1-\mathrm{N}^2|}}.
\end{equation}

There are in total three different processes that could lead to a fragmentation collision event: turbulent fragmentation, drift-induced fragmentation, and advection-driven fragmentation, as described by Eqs. \ref{eq:frag}, \ref{eq:driftfrag}, and \ref{eq:advfrag}. When computing the size limits, we take the minimum of these three limits as our maximum size limits.

    \begin{figure*}[ht]
         \centering
        \includegraphics{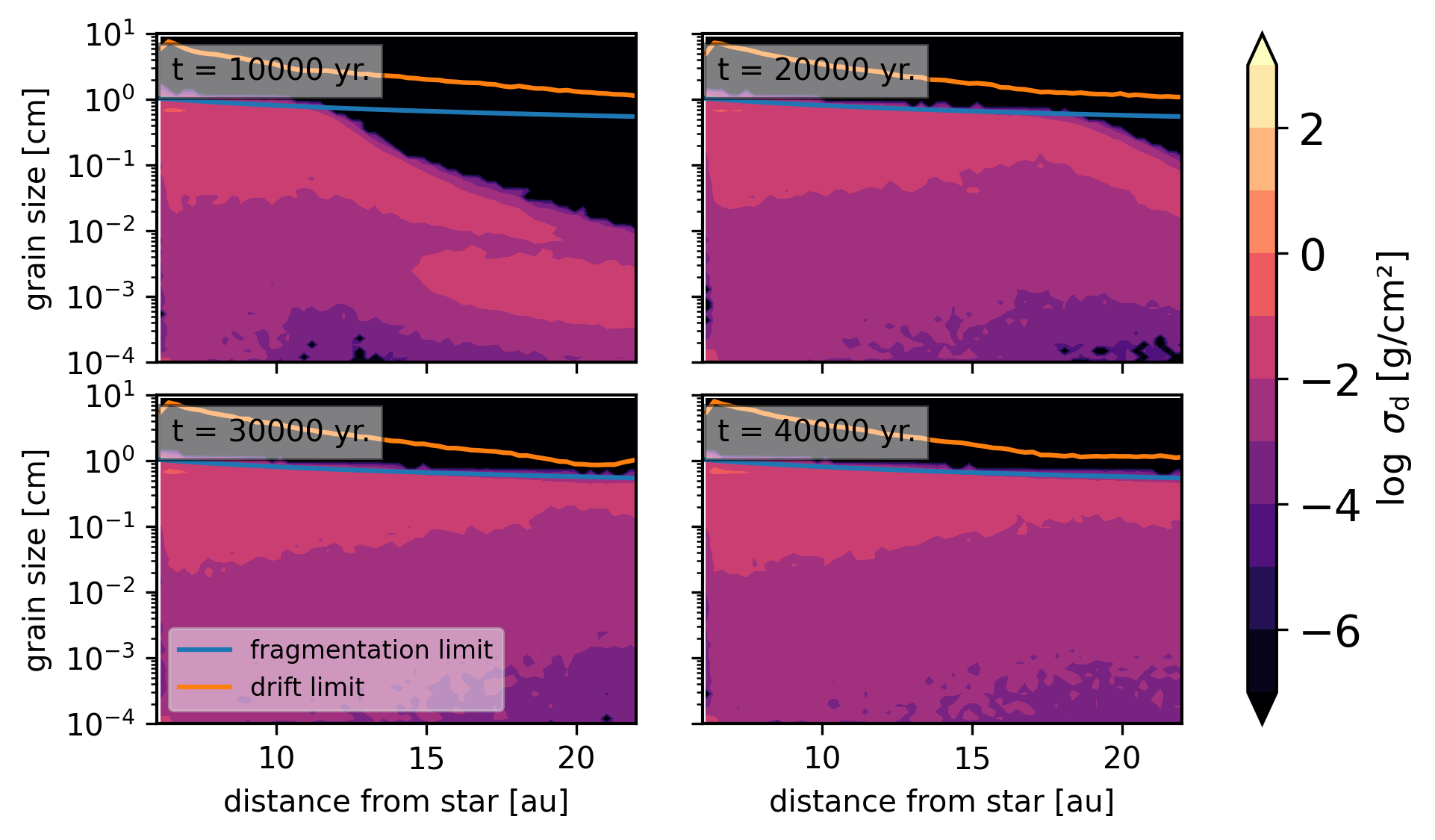}
        \caption{The time evolution of dust surface density $\sigma_{\rm d}$ as a function of grain size and distance from the central star for the \texttt{fidss} simulation with steady-state gas velocities. The blue and orange lines denote the fragmentation and drift limits (Eqs.~\ref{eq:frag} and~\ref{eq:drift}), respectively.}
        \label{fig:sigmada_fidss}
    \end{figure*} 

\begin{figure}[ht]
    \centering
    \includegraphics[scale=0.55]{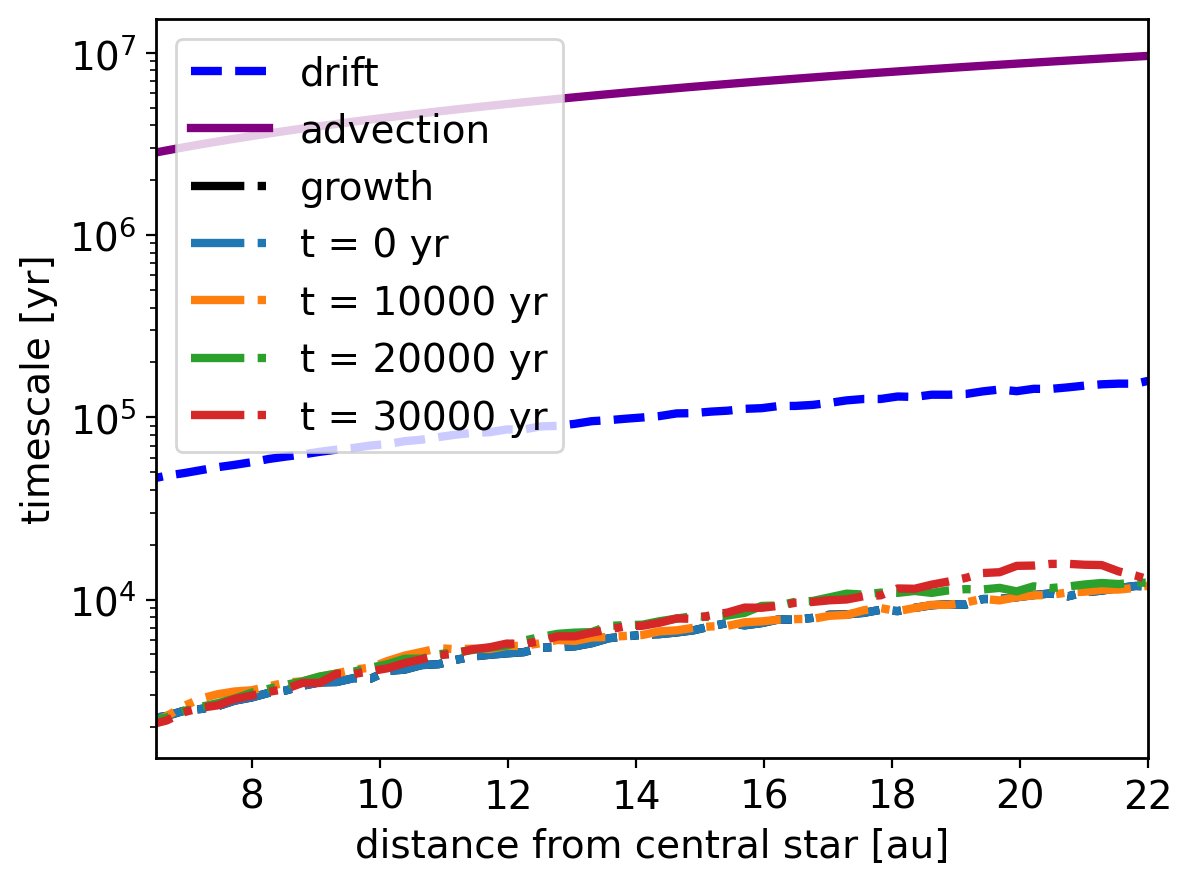}
    \caption{Time evolution of the growth timescale along with the drift and advection timescales for the \texttt{fidss} simulation with steady-state gas velocity.}
    \label{fig:fidsstimescale}
\end{figure}

\begin{figure}[ht]
    \centering
    \includegraphics[width=\linewidth]{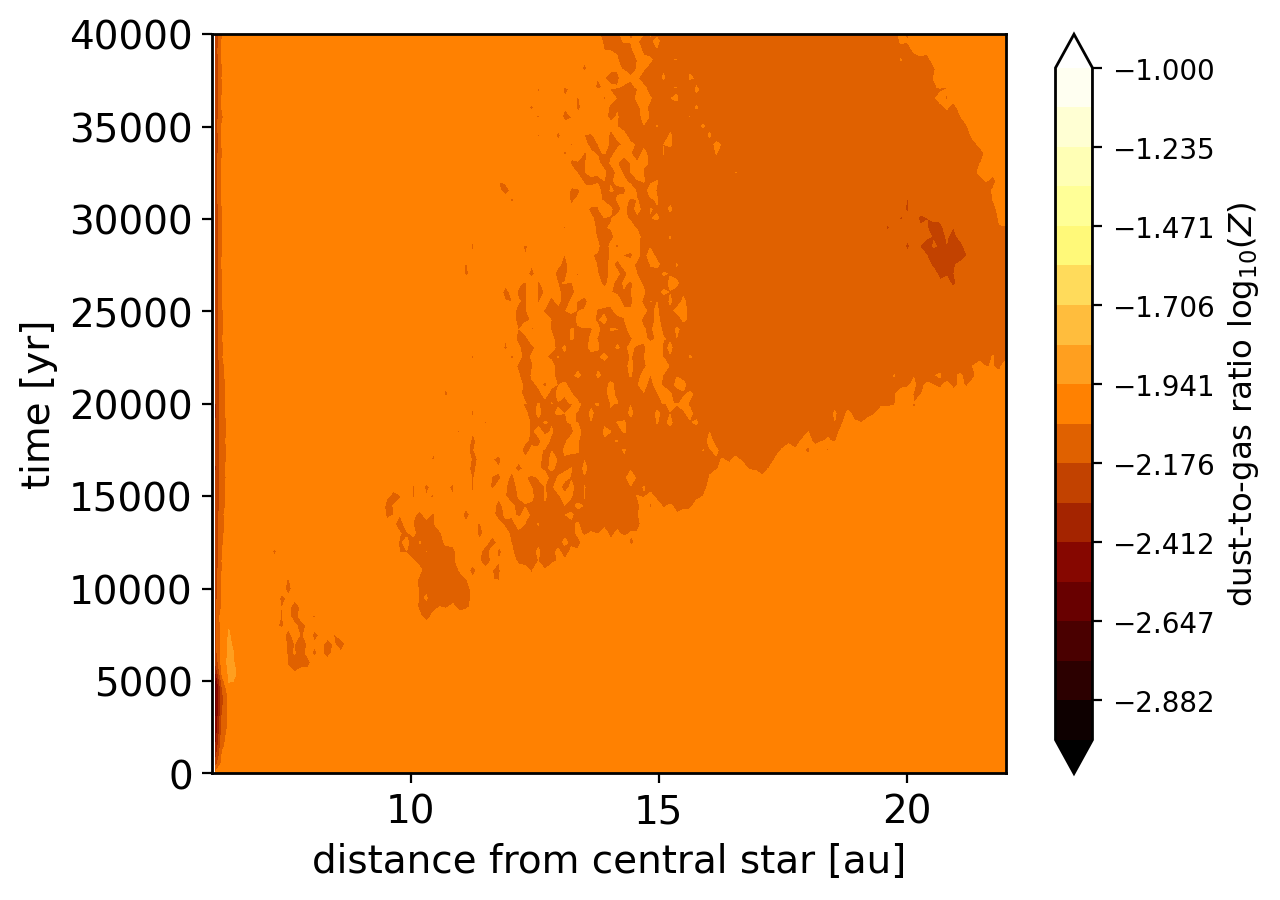}
    \caption{Evolution of the vertically integrated dust-to-gas ratios through time for the \texttt{fidss} simulation with steady-state gas velocity. }
    \label{fig:fidsssigmart}
\end{figure}

  \begin{figure*}[ht]
   \centering
   \includegraphics{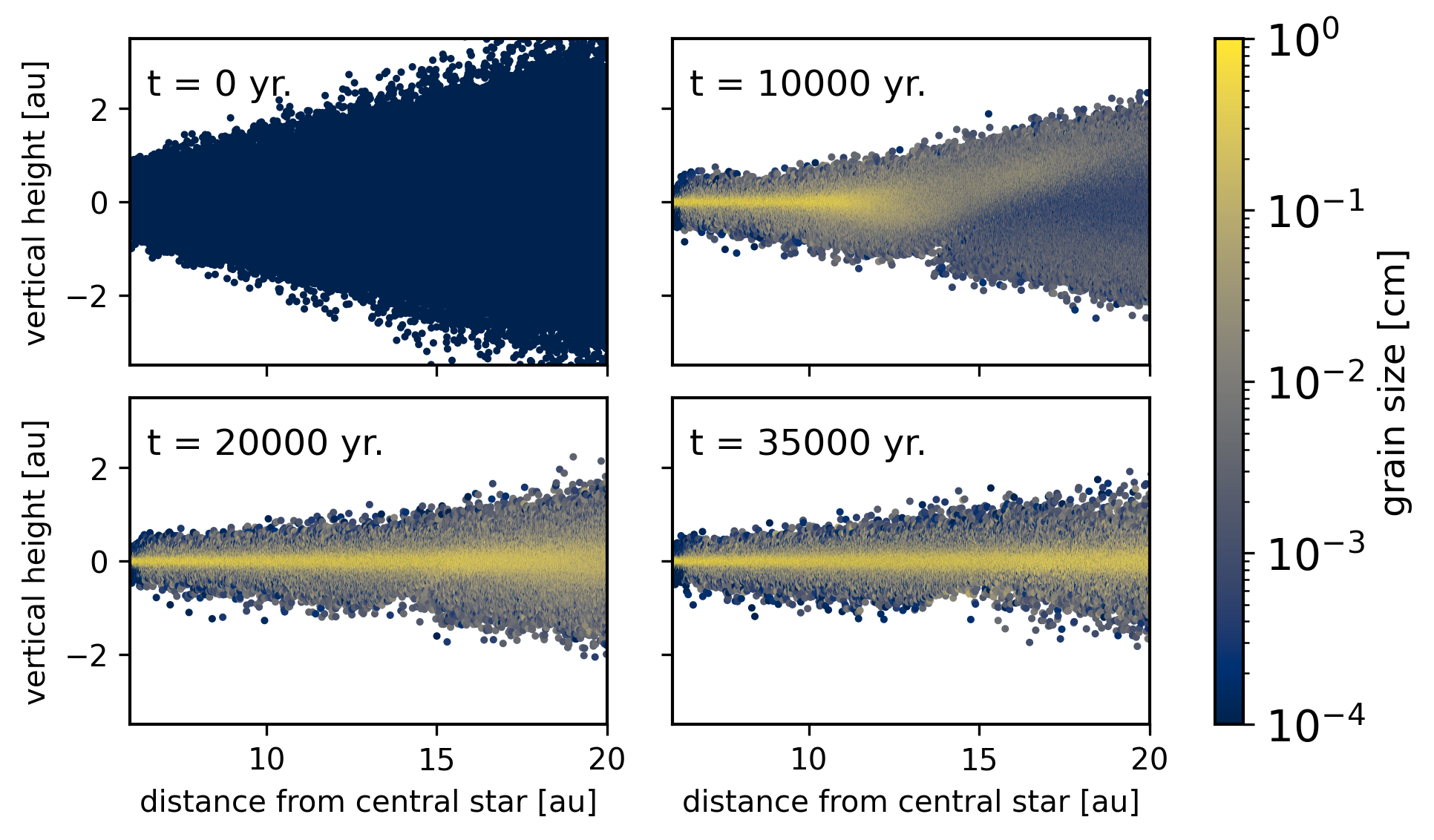}
   \caption{Same as Fig. \ref{fig:nowindscatter} but for the \texttt{fidmhd} simulation with the MHD gas velocities.}
              \label{fig:windscatter}
    \end{figure*}

    \begin{figure*}[ht]
   \centering
   \includegraphics{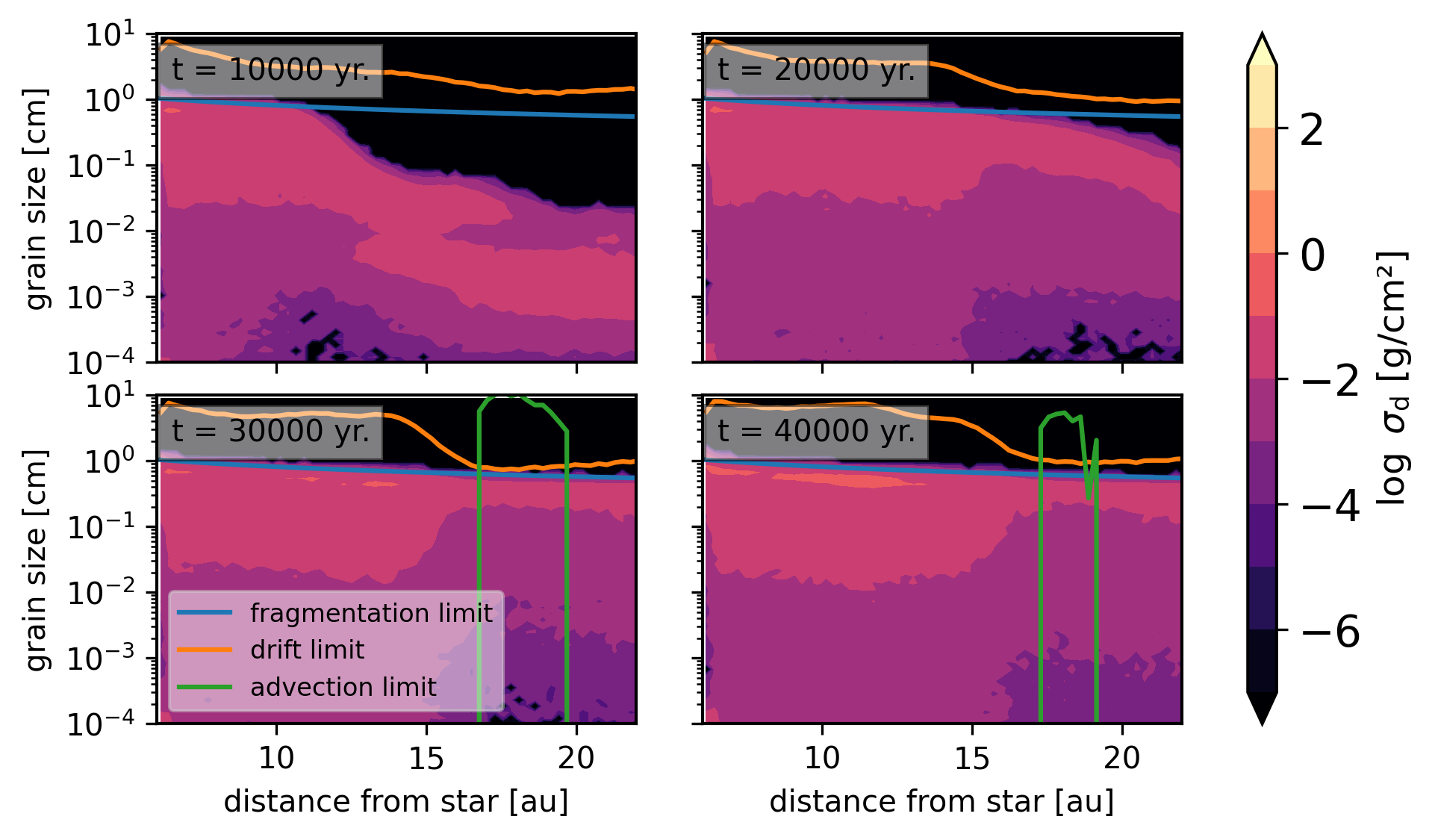}
   \caption{Same as Fig. \ref{fig:sigmada_fidss} but for the \texttt{fidmhd} simulation with the MHD gas velocities. Along with the fragmentation and the drift limits, we have the size limit of particles affected by gas advection(Eq. ~\ref{eq:adv}) plotted in green.}
              \label{fig:sigmada_fidmhd}
    \end{figure*}
\begin{figure}[ht]
    \centering
    \includegraphics[scale=0.55]{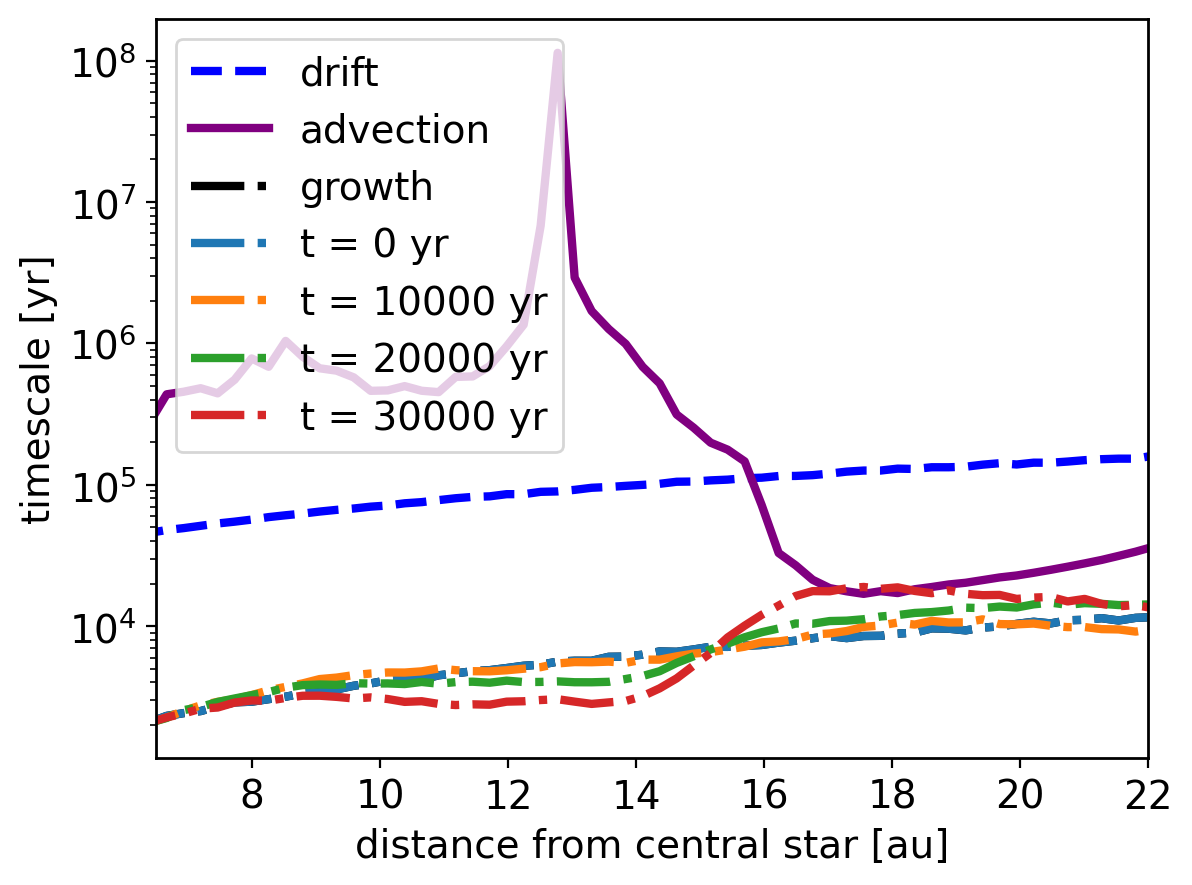}
    \caption{Same as Fig. \ref{fig:fidsstimescale} but for the \texttt{fidmhd} simulation with the MHD gas velocities. It can be seen that the advection timescale is comparable to the growth timescale at t=30000 yr.}
    \label{fig:fidmhdtimescale}
\end{figure}

\begin{figure}[ht]
    \centering
    \includegraphics[width=\linewidth]{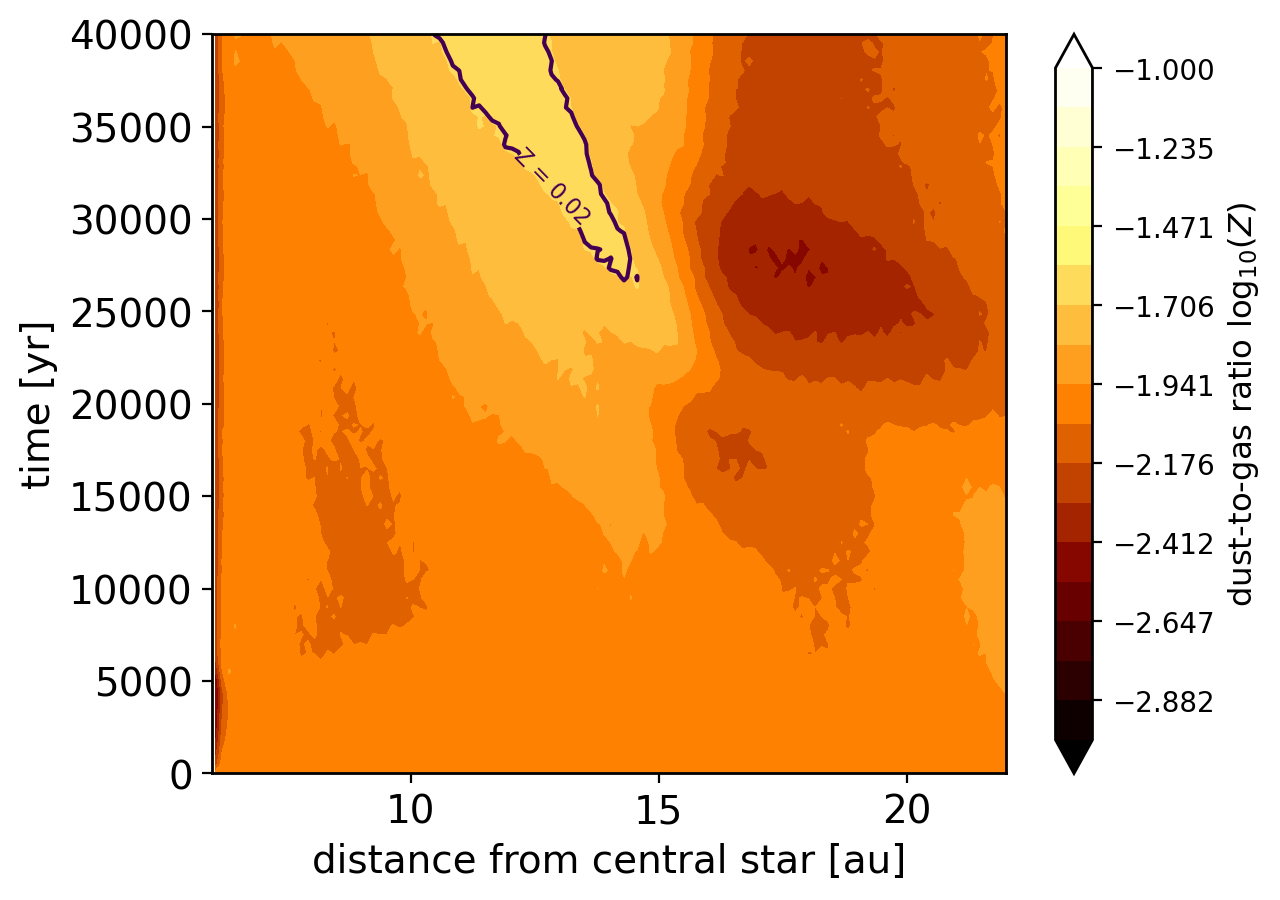}
    \caption{Same as Fig. \ref{fig:fidsssigmart} but for the \texttt{fidmhd} simulation with the MHD gas velocities. The Z=0.02 contour encloses the region in the parameter space where the dust-to-gas ratio is at least twice the initial value. }
    \label{fig:fidmhdsigmart}
\end{figure}

\begin{figure}
     \centering
     \includegraphics[width=\linewidth]{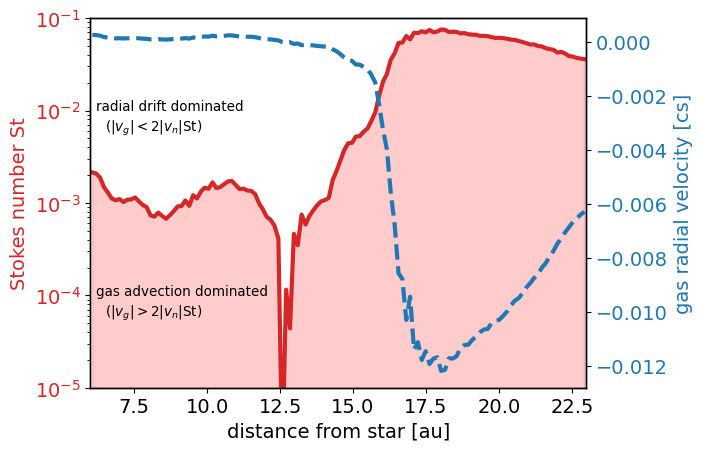}
     \caption{The red solid line denotes the Stokes number at which the gas velocity component of the dust radial velocity equals the radial drift component. We also plot the midplane gas radial velocity with the blue dashed line.}
     \label{fig:Stregime}
\end{figure}

\section{Results}\label{sec:results}
In this section, we outline the results of the different simulations as described in Sect. \ref{sec:simsetup} . First, we discuss the steady state simulation as a benchmark for dust size distributions in classical dust growth simulations. Next, we introduce the simulations with the gas velocities from B17 and compare them with the steady state simulations. We then discuss the results of varying the fragmentation velocity of the simulations and the implications for planetesimal formation. We further discuss the underlying processes that lead to the dust distributions for the simulations.

 \subsection{Steady state gas velocity}\label{subsec:fidss}
 We first analyse the simulation \texttt{fidss} where we use the steady-state gas velocity as described in Sec.~\ref{subsubsec:ssgas}.  Figure~\ref{fig:nowindscatter} shows the time evolution of the simulation. As mentioned in Sec.~\ref{sec:simsetup}, we start with micrometer-sized particles and let the system evolve. A notable feature of the simulation is the sedimentation-driven coagulation process as seen at t=10000 yr in Fig.~\ref{fig:nowindscatter}, where the growth is accelerated in the regions above and below the midplane. This happens due to the fact that the vertical settling velocity increases with height \citep{Drazkowska2013}. This leads to speeding up of particle growth in the upper regions, and these particles rain down to the midplane of the disk. This happens during the early periods of the simulation, and once the rain-out occurs, the largest particles are settled into the midplane and drift towards the inner disk. The simulation reaches a maximum dust size of $\sim 1.5$~cm in the fragmentation-limited inner region of the domain.

Figure \ref{fig:sigmada_fidss} shows the time evolution of the dust surface density as a function of grain size and radius overplotted with the fragmentation and drift size limits as described in Sec.~\ref{sec:diagnostics}. Turbulent fragmentation is the primary size-limiting process in the region, setting the particle sizes to $\lesssim 1$ cm. The simulation domain remains fragmentation-limited, meaning the growth timescale is shorter than the drift timescale, allowing the particles to reach the sizes where fragmentation occurs. The dust particles in inner areas reach their maximum sizes the earliest, as the growth timescale is shorter for smaller distances (see Eq. \ref{eq:growthtimescale2}) and there is an inside-out growth of particles in the disk.

This can also be seen by comparing the timescales of growth, drift, and advection as shown in Fig.~\ref{fig:fidsstimescale}. For a typical steady-state viscous disk, the advection timescales are too long to play an important role in setting the global processes. Also, the radial drift timescale is too long to affect the growth. In Fig.~\ref{fig:fidsssigmart}, we show the evolution of the vertically integrated dust-to-gas ratio through our domain. As mentioned in Sec.~\ref{sec:simsetup} and Table \ref{tab:fidparams}, we set up our simulation with a dust-to-gas ratio of 0.01 and as the system evolves the amount of solids in the outer disk reduces due to particles growing and drifting inwards resulting in a slowly reducing dust-to-gas ratio from the outer boundary to the inner boundary, but is refilled by the incoming feeding flux.


        \begin{figure*}
   \centering
   \includegraphics{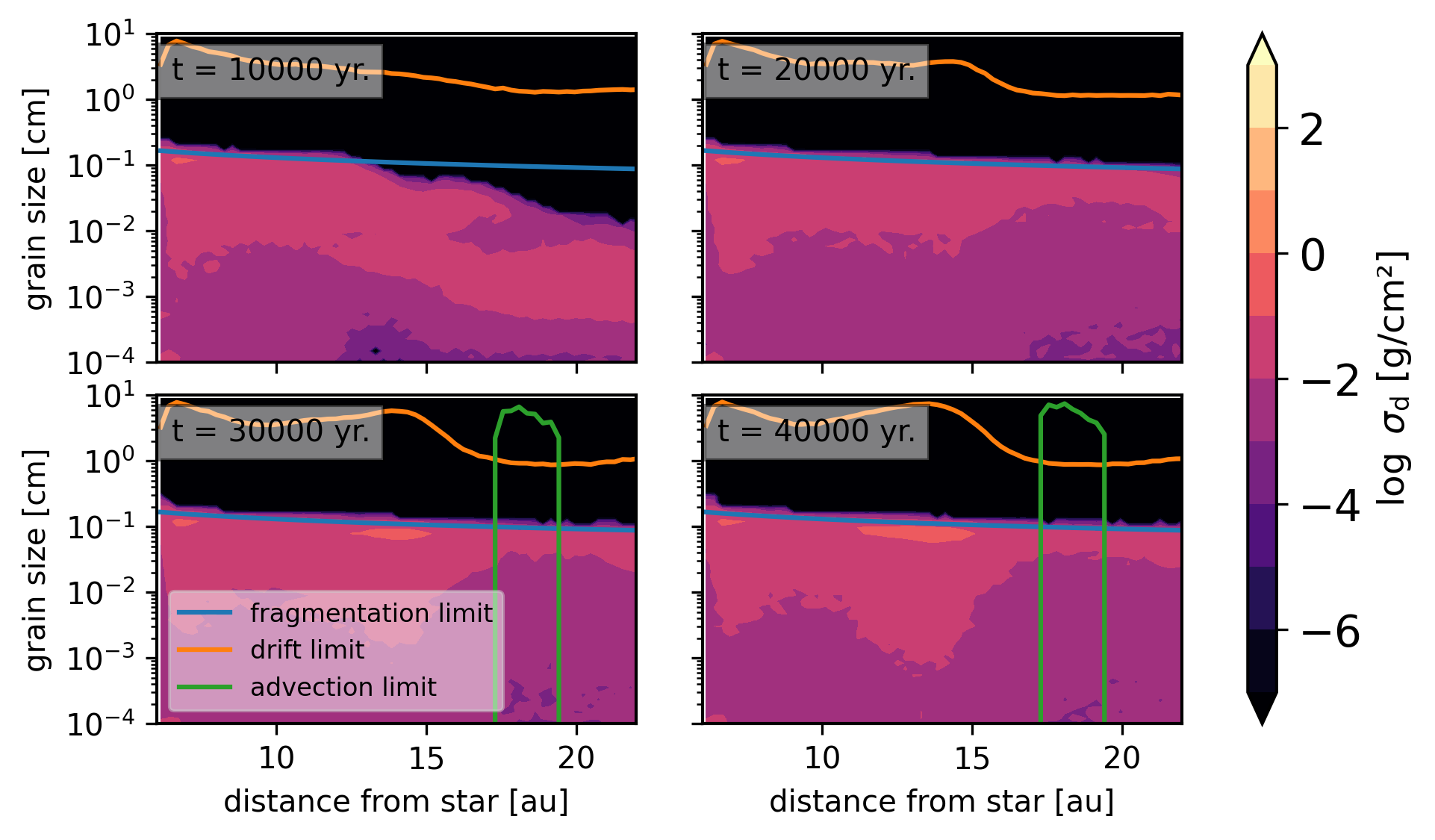}
 \caption{Same as Fig.~\ref{fig:sigmada_fidmhd} but for the simulation \texttt{mhdlowvfrag} with MHD gas velocities and $v_{\mathrm{frag}} =100 \mathrm{cm/s}$.}
    \label{fig:sigmada_mhdvfrag}
    \end{figure*}

\begin{figure}[ht]
    \centering
    \includegraphics[width=\linewidth]{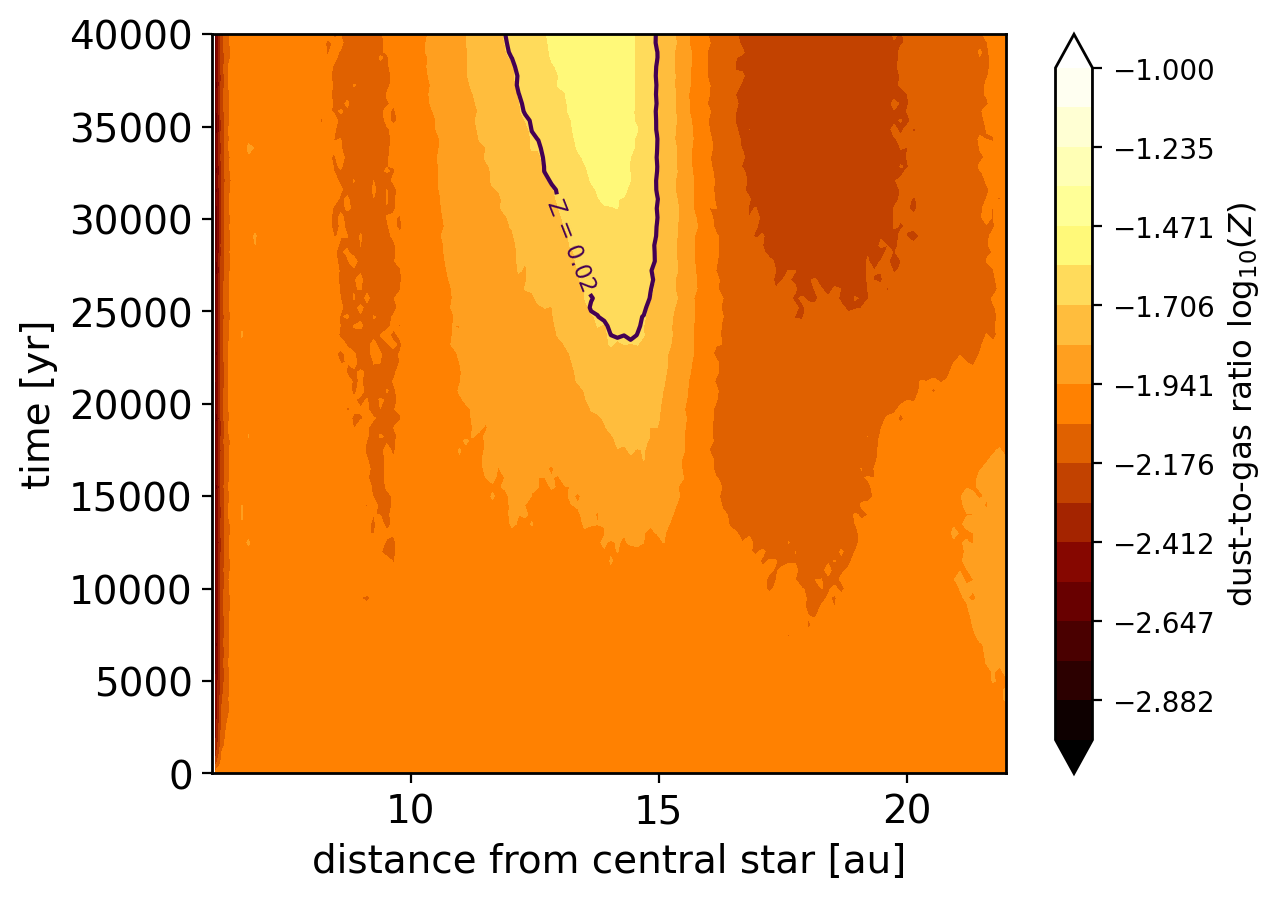}
    \caption{Same as Fig.~\ref{fig:fidmhdtimescale} but for the simulation \texttt{mhdlowvfrag} with MHD gas velocities and $v_{\mathrm{frag}} =100 \mathrm{cm/s}$.}
    \label{fig:mhdlowvfragsigmart}
\end{figure}


\subsection{MHD gas velocity}

We analyze the results of the simulation \texttt{fidmhd} where we use the gas velocity data from B17.  The gas flow architecture in the non-ideal MHD is such that the outer region has a symmetric gas flow with an enhanced midplane accretion flow that transitions at $\sim 15-17$ AU into an anti-symmetric flow as described in Sec.~\ref{subsubsec:mhdgas} and shown in Fig.~\ref{fig:windvel}. This architecture is expected to leave an imprint on the transport of dust as discussed in \citet{Hu2021}. Fig.~\ref{fig:windscatter} shows the time evolution of the model with the complex gas flow. Sedimentation-driven coagulation can be seen in the simulation at $t = 10000 $~yrs, but it is not similar to Fig.~\ref{fig:nowindscatter} as the asymmetries in the radial gas velocity (as seen in Fig.~\ref{fig:windvel}) impact the dust radial velocities and hence the collisional velocities of the particles, resulting in a different initial dust growth pattern. However, the particle maximum sizes are not affected, and remain similar to the \texttt{fidss} case.

Initially, dust growth happens in a two-pronged way where the inner disk grows to larger sizes faster due to the short growth timescales, but there is also an accelerated growth in the flow transition region at around 16 AU. This is evident in Fig.~\ref{fig:sigmada_fidmhd} at $t = 10000$~yrs. From $t=30000$~yrs, advection of the dust grains due to the gas velocity begins to affect the small particles. This can be seen by computing the different timescales. As seen in Fig.~\ref{fig:fidmhdtimescale}, it is the growth timescale that is the shortest throughout the domain initially, but as the system evolves, particles get transported inward/outward depending on their radial velocity, increasing the growth timescale in the outer part of the domain. A tipping point comes at $t=30000$~yrs, where the advection timescale and the growth timescale compete to be the shortest, meaning that these processes dominate the dust dynamics. These timescales set the size limits for the system as seen in Fig.~\ref{fig:sigmada_fidmhd}, where we plot the maximum midplane grain size for advection, fragmentation, and radial drift. Turbulent fragmentation sets the maximum size limits through the simulation. But in the regions between $\sim 16 -19~\mathrm{AU}$ of the disk, advection starts to play a dominant role. The midplane gas velocities are very strong, and this leads to almost all the particles in the region being impacted by advection. The small dust particles are the most affected by the strong gas velocities, and this is evident in the lower density of small particles in Fig.~\ref{fig:sigmada_fidmhd} as compared to Fig.~\ref{fig:sigmada_fidss}. We also computed drift-induced fragmentation and advection-induced fragmentation limits (Eqs.~\ref{eq:driftfrag} and~\ref{eq:advfrag}), but these size limits are not relevant to the dust size distributions. It is the competition between dust growth and advection that dominates in the outer part of our domain.

Figure \ref{fig:fidmhdsigmart} shows the dust-to-gas ratio evolution of the \texttt{fidmhd} simulation. There is an enhancement of solids that starts at around $t=10000$~yrs inside of 15~AU. The location of this bump corresponds to the transition zone of the gas flow from symmetric to anti-symmetric gas flow as seen in Fig.~\ref{fig:windvel}. The midplane gas velocities are getting lower as the particles drift to the inner part of the domain. This leads to a traffic-jam effect at the transition region as the particles that were moving inwards due to drift/advection now have a chance to collide with more particles. As the particles in the trap grow slightly faster, they drift inwards, moving the location of the dust overdensity. This overdensity in large particles can also be noticed in Fig. \ref{fig:sigmada_fidmhd} at $t=30000$~yrs at around 15 AU, but at $t=40000$~yrs the overdensity has already moved a few AU inwards. The bump reaches a maximum dust-to-gas ratio of 0.025, which is a moderate increase from our initial dust-to-gas ratio of 0.01.

To demonstrate the strength of the advection of dust particles due to the gas velocity, we analyze the two components of the radial velocity of dust, the advection of dust particles by gas (Eq. \ref{eq:advvelocity}) and the radial drift component (Eq.~\ref{eq:driftvel}). Figure \ref{fig:Stregime} shows the two regimes as a function of the Stokes number, where the gas advection component dominates over the radial drift component and vice versa. The red solid line denotes the Stokes number for which both the components are equal. For our MHD gas velocities from B17, we see that for the outer disk, the dominant contribution to the radial velocity for the dust particles is from the gas advection component.

\subsection{Lower fragmentation velocity}
We analyse the results of the simulation \texttt{mhdlowvfrag}, where we use the gas velocity data from B17, but here we reduce the fragmentation velocity to 100~cm/s as detailed in Table \ref{tab:listofsims}. Laboratory experiments of dust collisions suggest that the fragmentation velocity is closer to 100~cm/s \citep{Guettler2010}, motivating the parameter variation in this study. The underlying processes are similar to \texttt{fidmhd}, so the simulation starts similarly, but then due to the reduced fragmentation velocity, the particles do not get to grow to sizes as large as seen in \texttt{fidmhd} and \texttt{fidss}. The maximum particle size in the simulation is $\sim 0.3$~cm, which is an order of magnitude smaller than in the previous simulations.

Figure \ref{fig:sigmada_mhdvfrag} shows the surface density evolution. The initial two-pronged growth present in the \texttt{fidmhd} simulation at $t=10000$~yrs is seen in the \texttt{mhdlowvfrag} case as well. There is a higher density of smaller particles due to the lowering of the fragmentation velocity to 100 cm/s. More particles are affected by advection when compared to the \texttt{fidmhd} case and can remove small particles and influence the size distributions as seen at $t = 30000$~yr in Fig. \ref{fig:sigmada_mhdvfrag}. A particle overdensity at around 15 AU is noticeable, which corresponds to the flow transition region. Although this was seen in the \texttt{fidmhd} simulation, the difference is that the pile-up seen in Fig. \ref{fig:sigmada_mhdvfrag} for the \texttt{mhdlowvfrag} simulation does not drift inwards as fast as seen in the \texttt{fidmhd} case. This is also evident when looking at the vertically integrated dust-to-gas ratios as seen in Fig.~\ref{fig:mhdlowvfragsigmart}. The overdensity stays close to 15 AU from the time it is forming to the end of the simulation. By reducing the fragmentation velocity, we are able to trap the particles a bit more efficiently. Since most of the particles in the pile-up are large in size, they can have an effect on planetesimal formation. We discuss the effects in the following section.

\subsection{Planetesimal formation}\label{subsec:planetesimalformation}

We investigate the chances of forming planetesimals in the three simulations \texttt{fidmhd}, \texttt{fidss}, and \texttt{mhdlowvfrag}. In protoplanetary disks, the leading process that is known to form planetesimals is the streaming instability \citep[SI;][]{Johansen2007,Drazkowska2023}, a two-fluid instability that can lead to clumping of solids until they reach densities required to form planetesimals by gravitational collapse of those clumps. Recent studies have explored the conditions for forming the clumps and their subsequent collapse, and these can be condensed to the requirement of a midplane pebble-to-gas ratio of about unity \citep{Lim2024}. Typically, the larger the Stokes number of the particles that are being clumped, the easier to form the planetesimals. In our work, we define a pebble to be any particle with $\mathrm{St}> 10^{-3}$ in accordance with \citet{Birnstiel2024}, who define a pebble as any particle that has $\mathrm{St} > \alpha$. 

\begin{figure}
    \centering
    \includegraphics[width=\linewidth]{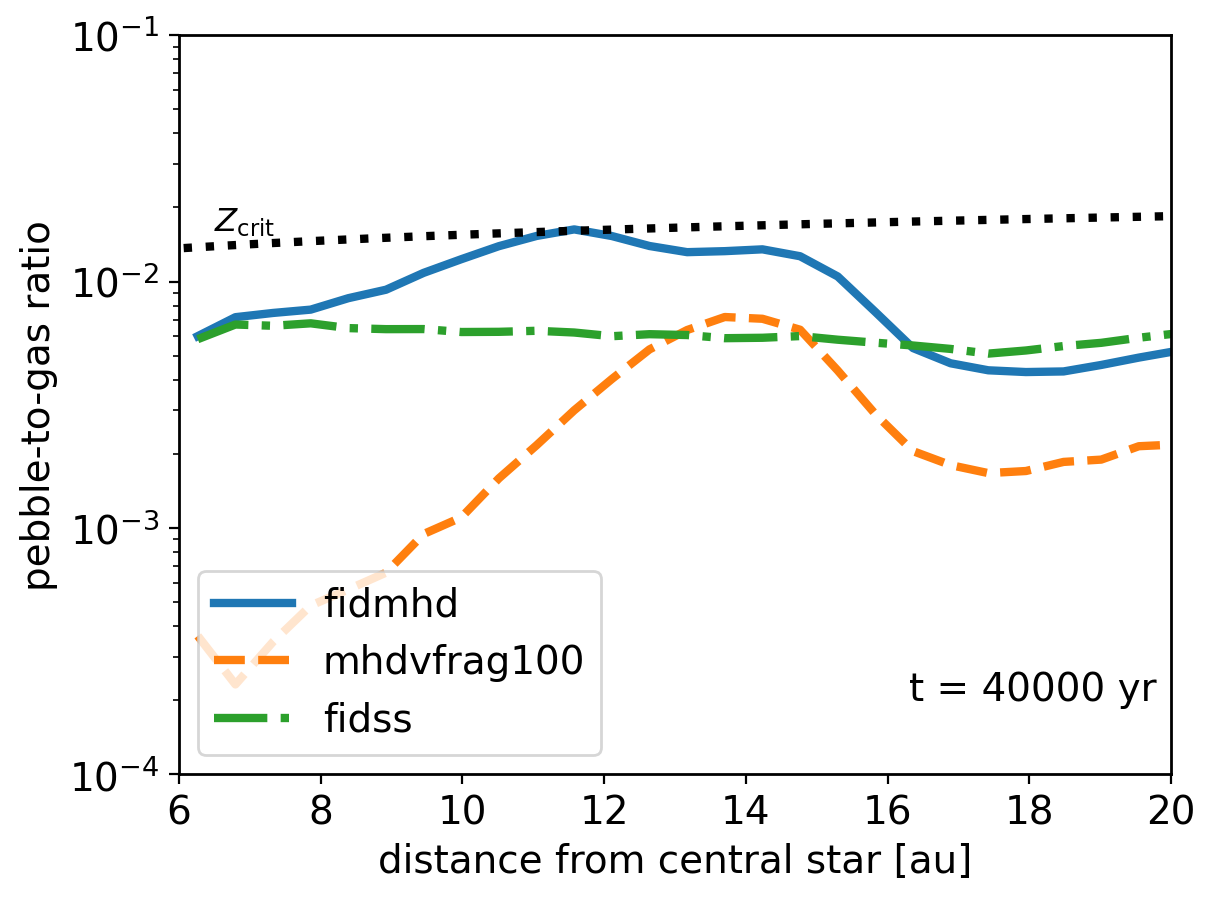}
    \caption{The vertically integrated pebble-to-gas ratios for the three simulations \texttt{fidmhd}, \texttt{mhdlowvfrag}, \texttt{fidss} at t = 30000 yr. Also plotted is the $Z_{\mathrm{crit}}$ to trigger SI as given by Eq. \ref{eq:Zcritscaled}.}
    \label{fig:pebble1D}
\end{figure}

Trying to compute the midplane pebble-to-gas ratios can be very resolution dependent, and to overcome the resolution issues, we look at the vertically integrated pebble-to-gas ratios for the three simulations. \citet{Lim2025} computed the critical vertically integrated pebble-to-gas ratios to trigger SI for smaller Stokes numbers and derived an analytical fit, which is given by
\begin{equation}\label{eq:Zcrit}
\mathrm{log}(Z'_{\mathrm{crit}}) = A (\mathrm{log}\mathrm{St})^2 + B\:\mathrm{log}\mathrm{St} + C,
\end{equation}
where A = 0.01, B = 0.07 and C = -2.36 are constants for the best fit of the parameters. The critical pebble-to-gas ratios depend on the radial gas pressure gradient \citep{Bai2010, Sekiya2018} at the location of the clump, which is given by
\begin{equation} \label{eq:gaspresgrad}
    \Pi = - \frac{1}{2} \left(\frac{H_g}{r}\right) \frac{\partial\; \mathrm{ln}\;P}{\partial \; \mathrm{ln} \;r} .
\end{equation}
\citet{Sekiya2018} note that the ratio $Z'_{\mathrm{crit}}/\Pi$ is more suited to be a universal parameter to investigate clumping by SI than $Z'_{\mathrm{crit}}$. \citet{Lim2025} used a value of $\Pi=0.05$ to compute the critical pebble-to-gas ratios in their simulations of SI and subsequently derive Eq.~\ref{eq:Zcrit}. So we rescale $Z'_{\mathrm{crit}}$ with $\Pi$ to arrive at the critical pebble-to-gas ratio given by,
\begin{equation}\label{eq:Zcritscaled}
    Z_{\mathrm{crit}} =\Pi \frac{Z'_{\mathrm{crit}}}{0.05}.
\end{equation}

Shown in Fig.~\ref{fig:pebble1D} are the vertically integrated pebble-to-gas ratios for the three simulations \texttt{fidss}, \texttt{fidmhd} and \texttt{mhdlowvfrag} at $t = 40000$~yr. We overplot the critical vertically integrated pebble-to-gas ratio for St $=3 \times 10^{-3}$ computed with the empirical relation for smaller Stokes numbers given by Eq.~\ref{eq:Zcrit} and rescaling it with Eq.~\ref{eq:Zcritscaled}. We note that Eq.~\ref{eq:Zcrit} has been computed for the case where there is no other source of turbulence (MRI, VSI, etc.) in their simulated domain other than the streaming instability-induced turbulence. Although we have turbulence in the form of parameterized $\alpha$ in our model, there is currently no other formula, so we use this as a reference and not as a hard criterion to trigger SI, since we do not include planetesimal formation in the models. For the simulations of SI with larger particles and external turbulence, the critical pebble-to-gas ratio seems to increase with the increasing turbulence \citep{Lim2024}. The \texttt{fidsss} simulations with steady-state gas velocities do not produce any dust overdensity. For the \texttt{fidmhd} simulation, the dust-to-gas ratio just crosses the $Z_{\mathrm{crit}}$. But in the case of \texttt{mhdlowvfrag}, the pebble enhancement is not high enough to cross the $Z_{\mathrm{crit}}$ value.

\section{Discussion}\label{sec:discussion}

\begin{figure*}
    \centering
    \includegraphics[width=0.5\linewidth]{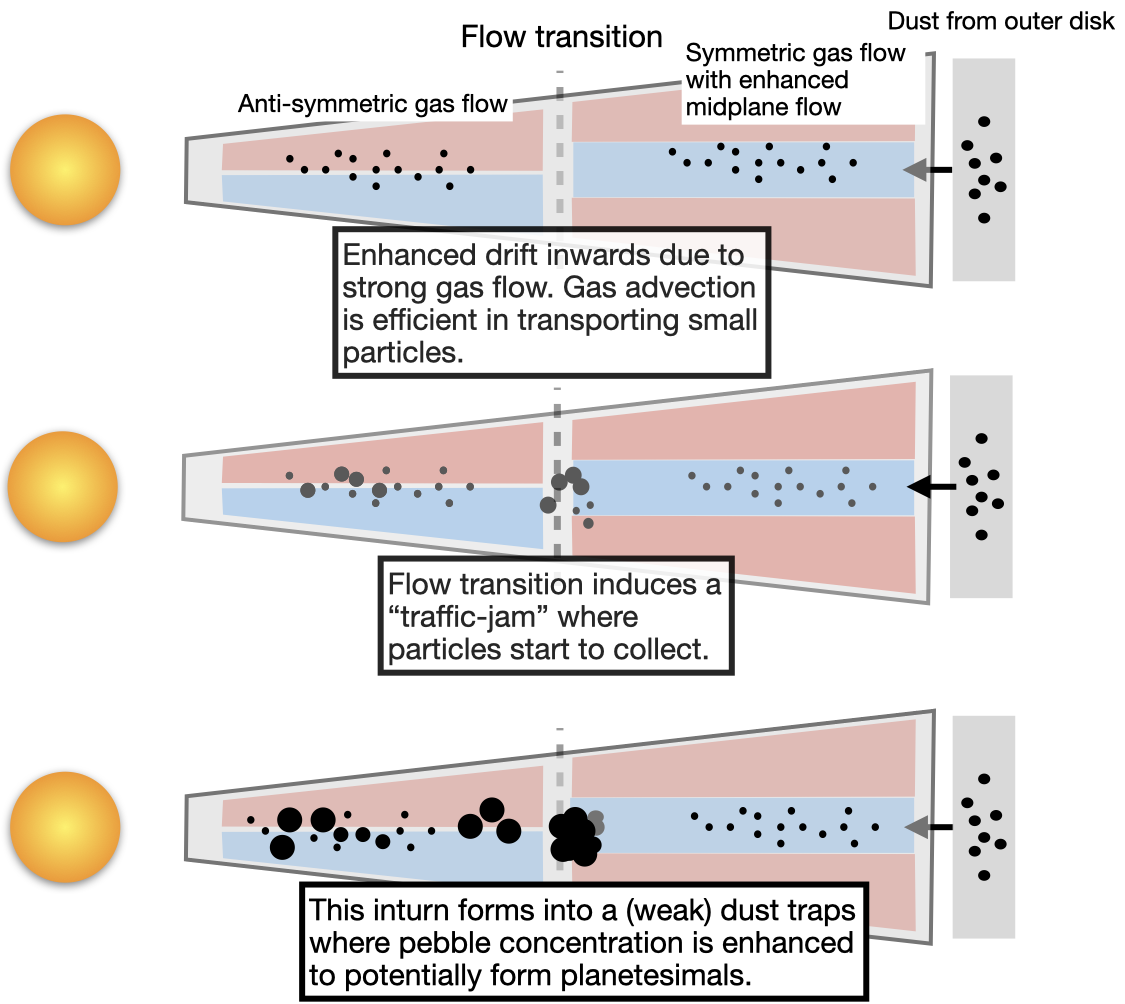}
    \caption{An overview of the sequence of processes that lead to the formation of the dust trap due to the complex gas flows in the simulations \texttt{fidmhd} and \texttt{mhdlowvfrag}.}
    \label{fig:summary}
\end{figure*}

\begin{figure}
\centering
\includegraphics[width=\linewidth]{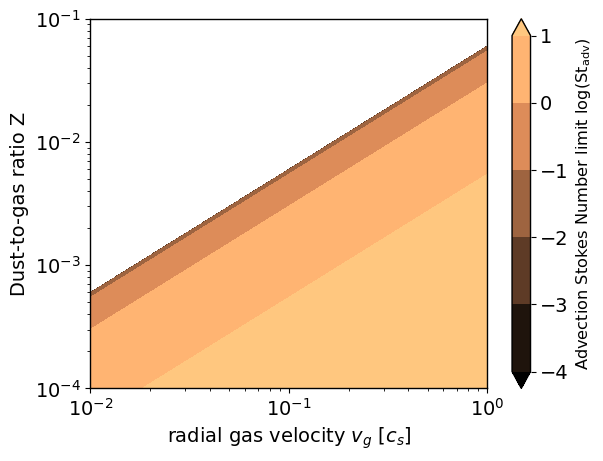}
\caption{The maximum Stokes number affected by advection as a function of dust-to-gas ratio and gas velocity.}
\label{fig:advectionlimits}
\end{figure}

In the previous section, we demonstrated that a complex gas flow can lead to the formation of a dust trap and enhanced pebble concentrations in this trap. Figure~\ref{fig:summary} shows a sketch summarising the sequence of processes that lead to the formation of the dust trap in magnetized disks with complex gas flows. The outer disk in our simulation has strong inward gas flows, which transport dust particles to the inner parts of the disk. But these dust particles face a traffic jam when they enter the regions of the disk where the midplane gas flows are not that strong anymore. This leads to a dust pile-up at this transition region, leading to enhanced concentrations of pebbles in the location.

\subsection{The influence of advection on dust size distribution}

Advection of dust particles due to gas flows plays an important role in forming such pile-ups, and this is an important process to be taken into account for gas flows of any origin, not just due to the Hall shear instability. Recent ALMA observations show strong gas flows \citep{Teague2019, Benisty2026}, and these flows can have an effect on the size distributions by reducing the amount of small dust. Such flows can also exist around a planet, as seen in simulations by \citet{Cilibrasi2023}, where the vertical velocities of gas can play a role in transporting the dust particles. The growth of particles will be inhibited because of advection depending on the magnitude of the gas velocity, local dust-to-gas ratio, and the radial location as seen in Eq.~\ref{eq:adv}. To demonstrate the importance of advection on dust size distribution, we compute the advection Stokes number $\mathrm{St}_{\mathrm{adv}}$ given by Eq.~\ref{eq:adv} for a range of dust-to-gas ratios and gas velocities at 20 AU. Figure~\ref{fig:advectionlimits} shows the maximum Stokes number that is influenced by advection for different dust-to-gas ratios and gas velocities. The advection growth barrier becomes more relevant with low dust-to-gas ratios and high gas velocities. At first glance, high gas velocities might not seem very common in disks, but \citet{Teague2019} found that the gas flows in HD163296 were on the order of 10\% sound speed, which, according to Fig.~\ref{fig:advectionlimits}, can be relevant to dust transport and growth already at dust-to-gas ratios $Z \sim 0.005$. Therefore, gas flows can play an important role in setting the dust size distributions based on their strength and the local environment in the disk.

\subsection{Forming dust traps without a pressure bump}

We found that the gas flow transition region could be a potential location for dust pile-up, particularly for lower fragmentation velocities. Recent observations have shown that dust in the protoplanetary disks may indeed be fragile \citep{Jiang2024}. Flow transitions are commonly seen in recent MHD simulations, and we propose that these transition regions may be a viable place for planetesimal formation. Typically, a pressure bump is invoked to trap dust and form planetesimals, and the pressure bumps are usually associated with an already existing planet. This poses a "chicken-and-egg" problem. If a planet is required to help form other planets, then we need to answer how the first planet formed without the help of a pressure bump created by a planet. Here, we demonstrate that gas flows with transition can pile up pebbles without having to invoke a pressure bump. 

In Sect.~\ref{subsec:planetesimalformation}, the simulations show an increase of pebble-to-gas ratio in the dust trap for both the simulations with the MHD gas velocities. Even though this may not directly fulfil the criteria for triggering SI and therefore forming planetesimals, this can act as the initial overdensity that is then needed to rapidly form structures that can efficiently form planetesimals. \citet{Carrera2025} showed that synergy between dust coagulation and SI can be very helpful in triggering SI starting from lower dust-to-gas ratios than previously shown by simulations. We do not model dust-to-gas feedback in our models, and this can become important in the dust trap, where such feedback processes slow down the drift and lower collisional velocities of dust particles and consequently aid in more particle clumping. All these factors together point that it may be possible to form planetesimals starting from the conditions we present here with our models.

The gas flows that we post-process are formed due to the HSI as discussed in Sect.~\ref{subsubsec:mhdgas}, and these processes are known to be important for the formation of the Solar System. Chondrules are millimeter-sized spherules found ubiquitously in chondritic meteorites,  making them an important aspect of understanding planet(esimal) formation in the Solar System. Chondrule formation requires localized heating events, and the large-scale gas flows due to HSI can produce current sheets that are ideal for such events \citep{Fu2023}. Although we do not investigate chondrule formation, having dust enhancements in the regions where the HSI is active can be a viable factor in aiding chondrule formation \citep{Alexander2008}.

\subsection{Caveats}

The gas flows have a lasting effect on the dust size distributions, but we do not explore the dynamic evolution of the gas flows. We use a time-averaged steady-state velocity background for our simulation. It is to be explored as to how long these structures persist. \citet{Hu2021} used the same data to explore dust transport with a similar model duration. However, more MHD simulations that explore further evolution of the disk are needed to establish the longevity of such flows. 

We also do not consider dust-to-gas backreaction and the action of dust (neutrals) on the chemical network and the magnetic field, which can potentially have effects on the non-ideal MHD \citep{Nolan2023, Okuzumi2019, Mori2019, XiaoHu2021}. The populations of neutrals (dust) are important to set the non-ideal MHD resistivities, especially for ambipolar diffusion. \citet{Wurster2021} found that the relative importance of the non-ideal MHD effects changes with different dust size distributions. \citet{Nolan2023} suggested that the zonal flows in gas rings that were typically observed in non-ideal MHD simulations could vanish when using certain static dust size distributions. The effect of dust growth, which is a dynamic process, is still to be explored in detail to know the actual effect on the non-ideal MHD effects and on the gas flow architecture. 

We perform all the simulations with the same turbulence parameter of $\alpha=3\times10^{-4}$ throughout the simulation domain. This value was chosen to be in agreement with the current inferred values of turbulence in disks \citep{Jiang2024, Villenave2025, Tong2026}. But there are disks where the trend varies \citep{Jiang2024}, for e.g., IM Lup has an $\alpha \sim 10^{-3}$ and HL Tau has a radially increasing $\alpha$. Modifying the strength of the turbulence in the disk could affect the efficiency of the dust traps. Stronger turbulence (increasing $\alpha$) would make dust transport due to turbulence stronger and hence weaken the dust trap more, and weaker turbulence (decreasing $\alpha$) would aid in more dust being trapped as dust transport due to turbulence is weakened. More investigation is required to understand the interplay of turbulence strength and gas advection on the dust size distributions.

\section{Summary and conclusions}  \label{sec:summary} 

We analysed the effect of a complex gas flow architecture from a non-ideal MHD simulation on the 2-D evolution of dust growth. We used gas velocity data from a non-ideal MHD simulation and compared the results to a standard model with steady-state gas velocity, which is typically used for a classical viscous disk. We also ran a simulation varying the fragmentation velocity to investigate its effect on the dust evolution in the presence of the complex gas flow. Finally, we investigate the effects of the flows on planetesimal formation. The major results are summarized here as follows:
\begin{itemize}
\item The complex gas flow architecture helps to form a dust trap in the flow transition region. The dust trap allows pebbles to concentrate, but once the pebbles grow large enough, the position of the ring moves inward.

\item Reducing the fragmentation velocity leads to a more stable ring as the smaller dust particles do not drift as fast and are more impacted by the gas advection flow. However, the lower fragmentation speed leads to smaller particles, and a lower dust-to-gas ratio is achieved in the trap.

\item In the presence of strong gas flows, the advection of dust particles plays a role in setting the dust size distributions in the disk. The transport of small dust is predominantly driven by the gas flows, and the population of the small particles in a given region of the disk is determined by the strength of the gas flows in that region. Removal of small dust particles from a region of strong gas flow may lead to a longer coagulation timescale.

\item The dust trap formed at the gas flow transition has a higher pebble-to-gas ratio compared to the rest of the disk. The dust density enhancement may be suitable for planetesimal formation without invoking a pressure bump, thereby helping to solve the chicken-and-egg problem of planet formation.
\end{itemize}

Overall, our results suggest that complex gas flows play a major role in setting the dust size distributions and potentially help us form planetesimals in the disk. With the rise in importance of magnetized disks, it is imperative that the effects of such magnetized disks on dust growth and planet formation are taken into account. 





\begin{acknowledgements}
     V.V thanks Alexandros Ziampras for insightful discussions. V.V. and J.D. were funded by the European Union under the European Union’s Horizon Europe Research \& Innovation Programme 101040037 (PLANETOIDS). Views and opinions expressed are, however, those of the author(s) only and do not necessarily reflect those of the European Union or the European Research Council. Neither the European Union nor the granting authority can be held responsible for them.
\end{acknowledgements}
\section{Data availability}
All data and code required to reproduce the results and figures from this paper are publicly available. The data can be accessed with the Zenodo repository at \href{https://doi.org/10.5281/zenodo.20274579}{https://doi.org/10.5281/zenodo.20274579}. The code to produce the results of the simulations and to reproduce the plots is available in the \href{https://github.com/vicky1997/dust_traps_complex_gas_flows_magnetized_protoplanetary_disks}{GitHub repository}. The base version of \texttt{mcdust} can be found in the GitHub repository, \href{https://github.com/vicky1997/mcdust}{https://github.com/vicky1997/mcdust} 

%
%

\bibliographystyle{aa}
\bibliography{bibliography}

\end{document}